\documentclass[11pt,twoside]{article}
\usepackage[top=1in, bottom=1in, left=1.25in, right=1.25in]{geometry}
\usepackage{latexsym}
\usepackage{amsmath}
\usepackage{amssymb}
\usepackage{txfonts}
\usepackage{graphicx}
\usepackage{epsfig}
\allowdisplaybreaks

\def\ra{\rangle}
\def\la{\langle}
\def\be{\begin{equation}}
\def\ee{\end{equation}}
\def\ba{\begin{array}}
\def\ea{\end{array}}

\begin{document}
\title
 {\Large \bf Entanglement detection and lower bound of
convex-roof extension of negativity}
\author
{ Ming Li$^1$\footnote{Tel.: +86-0532-86983375; E-mail address:
liming@upc.edu.cn.}, Tong-Jiang Yan$^{1,2}$ and Shao-Ming
Fei$^{3,4}$}
\date{}
\maketitle
\begin{center}
\begin{minipage}{140mm}

\small $~^{1}$ {\small College of Science, China University of
Petroleum, 266555, Qingdao}

{\small $~^{2}$ State Key Laboratory of Information Security
(Institute of Software, Chinese Academy of Sciences), 100049,
Beijing}

\small $~^{3}$ {\small School of Mathematical Sciences, Capital
Normal University, 100048 Beijing}

{\small $~^{4}$ Max-Planck-Institute for Mathematics in the
Sciences, 04103 Leipzig}

\vspace{2ex}

{\bf\small Abstract~} {\small We present a
set of inequalities based on mean values of quantum mechanical observables nonlinear entanglement witnesses for bipartite quantum
systems. These inequalities give rise to sufficient and necessary
conditions for separability of all bipartite pure states and even
some mixed states. In terms of these mean values of quantum mechanical observables a
measurable lower bound of the convex-roof extension of the negativity is derived.

{\bf Keywords~} Entanglement witness; Separability; Negativity.

\vfill }
\end{minipage}
\end{center}

\bigskip
Entanglement is not only the characteristic trait of quantum
mechanics, but also a vital resource for many aspects of quantum
information processing such as quantum computation, quantum
metrology, and quantum communication\cite{nielsen}. One of the
fundamental problems in quantum entanglement theory is to determine
which states are entangled and which are not, either theoretically
or experimentally. The entanglement witness \cite{wit1,wit2} is the
most useful approach to characterize quantum entanglement
experimentally. In recent years there have been considerable efforts
in constructing and analyzing the structure of entanglement witness
(see \cite{wit3,wit4,wit5,yusx,zhaomj} and the references therein).
Generally the Bell inequalities
\cite{bell,bell1,bell2,bell3,bell4,liprl} can be recast as
entanglement witnesses. Better entanglement witnesses can be also
constructed from more effective Bell-type inequalities.

On the other side, to quantify quantum entanglement is also a significant problem in quantum
information theory. A number of entanglement measures such as the
entanglement of formation and distillation \cite{Bennett96a, BBPS,
vedral02}, negativity \cite{Zyczkowski98a} and relative entropy
\cite{vedral02, sw} have been proposed for bipartite systems
\cite{BBPS} \cite{sw}-\cite{HillWootters}. The negativity was derived from the
positive partial transposition (PPT) \cite{ppt}. It bounds two relevant quantities characterizing the
entanglement of mixed states: the channel capacity and the
distillable entanglement. The convex-roof
extension of the negativity (CREN) \cite{dnk} gives a better characterization of entanglement,
which is nonzero for PPT entangled quantum states.

In this paper, similar to the non-linear entanglement witnesses and Bell-type inequalities,
we present a set of inequalities based on mean values of quantum mechanical observables, which
can serve as necessary and sufficient conditions for the separability of bipartite
pure quantum states and the isotropic states. These inequalities
are also closely related to the measure of quantum entanglement. According to the violation
of these inequalities, we derive an experimentally measurable
lower bound for the convex-roof extension of the negativity.

We first give a brief review of the 3-setting nonlinear entanglement
witnesses enforced by the indeterminacy relation of complementary
local observables for two-qubit systems \cite{yusx}. For a two-level system there are
three mutually complementary observables $A_i=\vec{a}_i\cdot
\vec{\sigma}$, where $\vec{a}_i$, $i=1,2,3$, are three normalized
vectors that are orthogonal to each other,
$\vec{\sigma}=(\sigma_x,\sigma_y,\sigma_z)$ are the Pauli matrices.
$\mu_A=-iA_1A_2A_3$ is the so called orientation of $A_i$s. $\mu_A$
takes values $\pm 1$. Similarly, one can define three mutually
complementary observables $B_i=\vec{b}_i\cdot \vec{\sigma}$ $(i=1;
2; 3)$ with the corresponding orientations $\mu_B$. It has been
shown that \cite{yusx}: (i) A 2-qubit state $\rho$ is separable if
and only if the following inequality holds for all sets of
observables $\{A_i, B_i\}_{i=1;2;3}$ with the same orientation:
\be\label{yuyu} \sqrt{\la A_1B_1+A_2B_2\ra^2_{\rho}+\la
A_3+B_3\ra^2_{\rho}}-\la A_3B_3\ra_{\rho}\leq 1; \ee (ii) For a
given entangled state the maximal violation of the above inequality
is $1-4\lambda_{\min}$, with $\lambda_{\min}$ being the minimal
eigenvalue of the partially transposed density matrix. The maximal
possible violation for all states is 3, which is attainable by the
maximal entangled states.

For qubit-qutrit systems, a similar inequality has been presented in
\cite{zhaomj}, which detects quantum entanglement also necessarily and sufficiently.
However the approaches in \cite{yusx} and \cite{zhaomj} can not be directly generalized to
higher dimensional systems, since it is based on the PPT criterion
that is both necessary and sufficient only for separability of
two-qubit and qubit-qutrit states. For general higher dimensional
$M\times N$ bipartite quantum systems a new approach has been
employed in \cite{liprl}. Let $\rho\in {\mathcal {H_{AB}}}$ be any
pure quantum states in vector space ${\mathcal {H_{AB}}}={\mathcal
{H_A}}\otimes {\mathcal {H_B}}$ with dimensions $dim\,{\mathcal
{H_A}}=M$ and $dim\,{\mathcal {H_B}}=N$ respectively. Assume
$L_\alpha^A$ and $L_\beta^B$ be the generators of special orthogonal
groups $SO(M)$ and $SO(N)$ respectively. The $M(M-1)/2$ generators
$L_\alpha^A$ are given by $\{|j\rangle\langle k|-|k\rangle\langle
j|\}$, $1\leq j < k \leq M$, where $|i\ra$, $i=1,...,M$, is the
usual canonical basis of ${\mathcal {H_A}}$, a column vector with
the $i$th row $1$ and the rest zeros. $L_\beta^B$ can be similarly
defined. The matrix operators $L_{\alpha}^A$ (resp. $L_{\beta}^B$)
have $M-2$ (resp. $N-2$) rows and $M-2$ (resp. $N-2$) columns that
are identically zero. We define the operators $A_i^{\alpha}$ (resp.
$B_j^{\beta}$) from $L_{\alpha}$ (resp. $L_{\beta}$) by replacing
the four entries in the positions of the two nonzero rows and two
nonzero columns of $L_{\alpha}$ (resp. $L_{\beta}$) with the
corresponding four entries of the matrix
$\vec{a_i}\cdot\vec{\sigma}$ (resp. $\vec{b_j}\cdot\vec{\sigma}$),
and keeping the other entries of $A_i^{\alpha}$ (resp.
$B_j^{\beta}$) zero.

By using $L_{\alpha}^A$ and $L_{\beta}^B$
the pure state $\rho$ can be projected to ``two-qubit" ones \cite{liprl}:
\begin{eqnarray}\label{qus}
\rho_{\alpha\beta}=\frac{L_{\alpha}^A\otimes
L_{\beta}^B\rho(L_{\alpha}^A)^{\dag}\otimes
(L_{\beta}^B)^{\dag}}{{\rm Tr}\{{L_{\alpha}^A\otimes
L_{\beta}^B\rho(L_{\alpha}^A)^{\dag}\otimes (L_{\beta}^B)^{\dag}}\}},
\end{eqnarray}
where $\alpha=1,2,\cdots, \frac{M(M-1)}{2}; \beta=1,2,\cdots,
\frac{N(N-1)}{2}$.
As the matrix $L_{\alpha}^A\otimes L_{\beta}^B$ has $MN-4$ rows and
$MN-4$ columns that are identically zero, one can directly verify
that there are at most $4\times4 = 16$ nonzero elements in each
matrix $\rho_{\alpha\beta}$.
For every pure state $\rho_{\alpha\beta}$ the
corresponding Bell operators are defined by
 \be\label{q} {\mathcal
{B}}_{\alpha\beta}=\tilde{A}_1^{\alpha}\otimes
\tilde{B}_1^{\beta}+\tilde{A}_1^{\alpha}\otimes
\tilde{B}_2^{\beta}+\tilde{A}_2^{\alpha}\otimes
\tilde{B}_1^{\beta}-\tilde{A}_2^{\alpha}\otimes \tilde{B}_2^{\beta},
\ee where
$\tilde{A}_i^{\alpha}=L_{\alpha}^AA_{i}^{\alpha}(L_{\alpha}^A)^{\dag}$
and $\tilde{B}_j^{\beta}=L_{\beta}^BB_j^{\beta}(L_{\beta}^B)^{\dag}$
are Hermitian operators. It has been shown that any
bipartite pure quantum state is entangled if and only if at least
one of the following inequalities is violated \cite{liprl},
\be\label{ob}
|\la{\mathcal {B}}_{\alpha\beta}\ra|\leq 2.
\ee

Inequalities (\ref{ob}) work only for general high dimensional
bipartite pure states. Combining the approaches in \cite{yusx} and
\cite{liprl}, we now define the mean value of nonlinear operators
${\mathcal {B}}^{'}_{\alpha\beta}$, \be \la{\mathcal
{B}}^{'}_{\alpha\beta}\ra=\sqrt{\la
\tilde{A}_1^{\alpha}\tilde{B}_1^{\beta}+\tilde{A}_2^{\alpha}\tilde{B}_2^{\beta}\ra^2_{\rho}+\la
\tilde{A}_3^{\alpha}+\tilde{B}_3^{\beta}\ra^2_{\rho}}-\la
\tilde{A}_3^{\alpha}\tilde{B}_3^{\beta}\ra_{\rho}, \ee for high
dimensional bipartite mixed states.

{\bf{Theorem 1:}} Any bipartite quantum state $\rho\in {\mathcal
{H_{AB}}}$ is entangled if any one of the following inequalities, \be
\label{nb} \frac{1}{{\rm Tr}(L_{\alpha}\otimes
L_{\beta}\,\rho^{T_A}L_{\alpha}\otimes L_{\beta})}|\la{\mathcal
{B}}^{'}_{\alpha\beta}\ra|\leq 1, \ee is violated, where
$\alpha=1,2,\cdots, \frac{M(M-1)}{2}$, $\beta=1,2,\cdots,
\frac{N(N-1)}{2}$.

{\bf{Proof:}} Assume that $\rho$ is separable (not entangled)
quantum state. Since the separability of a state does not change
under the local operation $L_{\alpha_0}^A\otimes L_{\beta_0}^B$, one
has that for any $\alpha$ and $\beta$,
$\rho_{\alpha\beta}=\frac{L_{\alpha}^A\otimes
L_{\beta}^B\rho(L_{\alpha}^A)^{\dag}\otimes
(L_{\beta}^B)^{\dag}}{{\rm Tr}\{{L_{\alpha}^A\otimes
L_{\beta}^B\rho(L_{\alpha}^A)^{\dag}\otimes
(L_{\beta}^B)^{\dag}}\}}$, which can be treated as a two qubits
state, must be also separable. According to the analysis in
\cite{yusx}, a 2-qubit state $\rho$ is separable if and only if
(\ref{yuyu}) holds, which contradicts with the condition (\ref{nb}).
Thus we have that if any one of the inequalities (\ref{nb}) is
violated, $\rho$ must be an entangled quantum state. $\hfill\Box$

It is obvious that the inequalities (\ref{nb}) must not be weaker
than the Bell inequalities given in \cite{liprl} for detecting
entanglement of mixed quantum states, since (\ref{nb}) supplies a
sufficient and necessary condition for separability of two qubits
(mixed) quantum states, while violating the CHSH inequality is just
a sufficient condition for two-qubit entanglement. Actually,
(\ref{nb}) is strictly stronger, as seen from the following
examples.

{\bf{Example 1}} We consider a $3\times 3$ dimensional state
introduced in \cite{bennett} by Bennett et al. Set $|\xi_{0}\ra =
\frac{1}{\sqrt{2}}|0\ra(|0\ra-|1\ra)$, $|\xi_{1}\ra =
\frac{1}{\sqrt{2}}(|0\ra-|1\ra)|2\ra$, $|\xi_{2}\ra =
\frac{1}{\sqrt{2}}|2\ra(|1\ra-|2\ra)$, $|\xi_{3}\ra =
\frac{1}{\sqrt{2}}(|1\ra-|2\ra)|0\ra$, $|\xi_{4}\ra =
\frac{1}{3}(|0\ra+|1\ra+|2\ra)(|0\ra+|1\ra+|2\ra)$. Let
\begin{eqnarray*}
\rho=\frac{1}{4}(I_{9}-\sum\limits_{i=0}^{4}|\xi_{i}\ra\la\xi_{i}|).
\end{eqnarray*}
This state is entangled according to the realignment criterion
\cite{chenkai}. We consider the mixture of $\rho$ and  the maximal
entangled singlet $P_+=|\psi_+\ra\la \psi_+|$, where
$|\psi_+\ra=\frac{1}{\sqrt{3}}\sum_{i=0}^{2}|ii\ra$:
\begin{eqnarray}
\rho_{p}=(1-p)\rho+p P_+.
\end{eqnarray}
By straightforward computation, the bell inequalities (\ref{ob})
detect entanglement for $0.57602\leq p \leq 1$, while (\ref{nb}) detect entanglement for $0.18221\leq p
\leq 1$.

{\bf{Example 2}} Consider the state
\begin{eqnarray}
\rho_{p}(a)=(1-p)\rho(a)+pP_+,
\end{eqnarray}
where
$$
\rho(a)=\frac{1}{8a+1}\left(%
    \begin{array}{ccccccccc}
      a & 0 & 0 & 0 & a & 0 & 0 & 0 & a\\
      0 & a & 0 & 0 & 0 & 0 & 0 & 0 & 0\\
      0 & 0 & a & 0 & 0 & 0 & 0 & 0 & 0\\
      0 & 0 & 0 & a & 0 & 0 & 0 & 0 & 0\\
      a & 0 & 0 & 0 & a & 0 & 0 & 0 & a\\
      0 & 0 & 0 & 0 & 0 & a & 0 & 0 & 0\\
      0 & 0 & 0 & 0 & 0 & 0 & \frac{1+a}{2} & 0 & \frac{\sqrt{1-a^{2}}}{2}\\
      0 & 0 & 0 & 0 & 0 & 0 & 0 & a & 0\\
      a & 0 & 0 & 0 & a & 0 & \frac{\sqrt{1-a^{2}}}{2} & 0 & \frac{1+a}{2}
    \end{array}\right),
    $$
is the weakly inseparable state given in \cite{pla232}, $0<a<1$.

\begin{figure}[h]
\begin{center}
\resizebox{10cm}{!}{\includegraphics{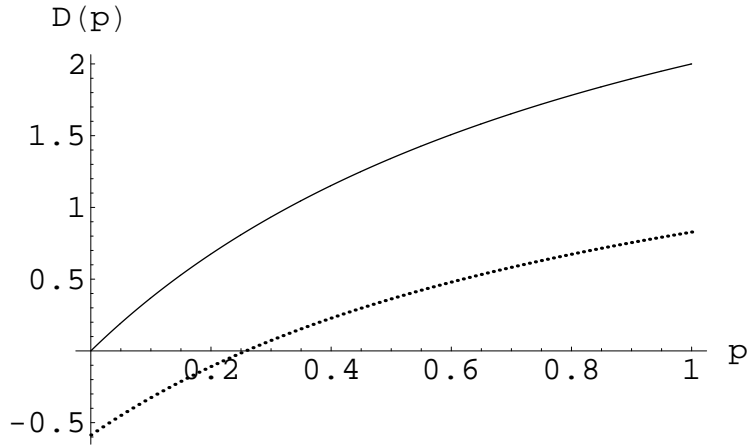}}
\end{center}
\caption{\label{fig1} The differences $D(p)$ between the right and
the left sides of the inequalities (\ref{nb}) (solid line) and the Bell
inequalities (\ref{ob}) (doted line).}
\end{figure}

Take $a=0.236$, which is the case that $\rho(a)$ violates the realignment
criterion \cite{chenkai} maximally. From Fig.\ref{fig1} we see that
the bell inequalities (\ref{ob}) detect entanglement for
$0.26\leq p \leq 1$, while (\ref{nb}) detect entanglement for the
whole region of $0< p\leq1$.

{\bf{Example 3}} Isotropic states \cite{isotro} with dimensions
$M=N=d$ can be written as the mixtures of the maximally mixed state
and the maximally entangled state
$|\psi_+\ra=\frac{1}{\sqrt{d}}\sum_{i=0}^{d-1}|ii\ra$,
\be\label{ist} \rho=\frac{1-x}{d^2}I_d \otimes
I_d+x|\psi_+\ra\la\psi_+|. \ee The inequalities (\ref{nb}) can
detect the entanglement for $x\leq \frac{1}{d+1}$ which agrees with
the result in \cite{isotro}. Thus (\ref{nb}) serves as a sufficient
and necessary condition of separability for isotropic states.

The inequalities (\ref{nb}) not only can be used to detect entanglement, but
also have some direct relations with the negativity. The negativity
of a bipartite quantum states $\rho$ with dimensions $d(H_A)=M$ and
$d(H_B)=N$ ($M\leq N$) is defined by \cite{dn} \be\label{d1}
{\mathcal {N}}(\rho)=\frac{||\rho^{T_A}||-1}{M-1},\ee where
$\rho^{T_A}$ is the partial transpose of $\rho$ and $||R||={\rm
Tr}\sqrt{RR^{\dag}}$ stands for the trace norm of matrix $R$. The
negativity is defined based on the positive partial transpose
criterion (PPT) \cite{ppt} which can not detect the PPT bound
entanglement. Thus it is not sufficient for the negativity to be a
good measure of entanglement. Lee et al in \cite{dnk} introduced the
convex-roof extension of the negativity (CREN) ${\mathcal
{N}}_m(\rho)$. For pure bipartite quantum states $|\psi\ra$,
${\mathcal {N}}_m(|\psi\ra)$ is exactly the negativity
 ${\mathcal {N}}(|\psi\ra)$ defined in (\ref{d1}). For a mixed bipartite quantum
 state $\rho$ the CREN is defined by
 \be\label{optim} {\mathcal {N}}_m(\rho)=
 \min\sum_kp_k{\mathcal {N}}_m(|\psi_k\ra),\ee
where the minimum is taken over all the ensemble decompositions of
$\rho=\sum_kp_k|\psi_k\ra\la\psi_k|$.

 The CREN can detect the PPT
bound entanglement, since it is zero if and only if the
corresponding quantum state is separable. Lee et al also show that
${\mathcal {N}}_m(\rho)$ does not increase under local quantum
operations and classical communication. However, generally it is
very difficult to calculate CREN analytically. Here we present an
experimentally measurable tight lower bound of CREN for arbitrary
bipartite quantum states, in terms of the violation of the
inequalities (\ref{nb}).

{\bf{Theorem 2:}} For any bipartite quantum states
$\rho\in{\mathcal{H_{AB}}}$, \be\label{lob} {\mathcal
{N}}_m(\rho)\geq\frac{1}{M-1}\sum_{\alpha\beta}|C_{\alpha\beta}|\,(\frac{X(\rho_{\alpha\beta})}{2}+1)-(M-1),
\ee where $C_{\alpha\beta}={\rm Tr}(L_{\alpha}\otimes
L_{\beta}\,\rho^{T_A}L_{\alpha}\otimes L_{\beta})$,
$X(\rho_{\alpha\beta})=\min\{0,d(\rho_{\alpha\beta})\}$, and
$d(\rho_{\alpha\beta})=\frac{1}{{\rm Tr}(L_{\alpha}\otimes
L_{\beta}\,\rho^{T_A}L_{\alpha}\otimes L_{\beta})}|\la{\mathcal
{B}}^{'}_{\alpha\beta}\ra|-1$ stands for the difference of the left
and right side of the inequalities (\ref{nb}).

{\bf{Proof:}} Let $|\psi\ra=\sum_i\sqrt{\mu_i}|ii\ra$ be a bipartite
pure state in Schmidt form. One has \be\label{p1} {\mathcal
{N}}_m(|\psi\ra)=\frac{2}{M-1}\sum_{i<j}\sqrt{\mu_i\mu_j}. \ee Note
that $\sum_i\mu_i=1$. By calculating the trace norm of
$L_{\alpha}\otimes
L_{\beta}(|\psi\ra\la\psi|)^{T_A}L_{\alpha}\otimes L_{\beta}$ for
each $\alpha$ and $\beta$, we derive that \be\label{p2}
\sum_{\alpha\beta}||C_{\alpha\beta}^{|\psi\ra}\, (|\psi\ra_{\alpha
\beta}\la\psi|)^{_{T_A}}||=(M-1)^2+2\sum_{i<j}\sqrt{\mu_i\mu_j}, \ee
where $|\psi\ra_{\alpha \beta}=\frac{L_{\alpha}\otimes
L_{\beta}|\psi\ra}{\sqrt{C_{\alpha\beta}^{|\psi\ra}}}$ and
$C_{\alpha\beta}^{|\psi\ra}={\rm Tr}\{L_{\alpha}\otimes
L_{\beta}|\psi\ra\la\psi|L_{\alpha}\otimes L_{\beta}\}$.

Let $\rho=\sum_kp_k\rho_k=\sum_kp_k|\psi_k\ra\la\psi_k|$ be the
optimal decomposition which fulfills that ${\mathcal {N}}_m(\rho)$
attains its minimum. In terms of (\ref{p1}) and (\ref{p2}) we get
\begin{eqnarray*}
{\mathcal {N}}_m(\rho)&=&\sum_kp_kN(\rho_k)\\
&=&\frac{1}{M-1}\sum_kp_k\sum_{\alpha\beta}||C_{\alpha\beta}^k\,(\rho_{\alpha\beta}^k)^{T_A}||-(M-1)\\
&\geq&\frac{1}{M-1}\sum_{\alpha\beta}||\sum_k p_k\, C_{\alpha\beta}^k\, (\rho_{\alpha\beta}^k)^{T_A}||-(M-1)\\
&=&\frac{1}{M-1}\sum_{\alpha\beta}||\sum_k p_k\, L_{\alpha}\otimes
L_{\beta}\rho_k^{T_A} L_{\alpha}\otimes L_{\beta}||-(M-1)\\
&=&\frac{1}{M-1}\sum_{\alpha\beta}||L_{\alpha}\otimes
L_{\beta}\rho^{T_A} L_{\alpha}\otimes L_{\beta}||-(M-1)\\
&=&\frac{1}{M-1}\sum_{\alpha\beta}|C_{\alpha\beta}|\,||\rho_{\alpha\beta}^{T_A}||-(M-1)\\
&=&\frac{1}{M-1}\sum_{\alpha\beta}|C_{\alpha\beta}|(\frac{X(\rho_{\alpha\beta})}{2}+1)-(M-1),
\end{eqnarray*}
where we have used that $||\rho_{\alpha\beta}^{T_A}||$ has at most one
negative eigenvalue (see \cite{verstraete}) in deriving the last equation. $\hfill\Box$

{\bf{Remark:}} For the isotropic states (\ref{ist}) our lower bound
(\ref{lob}) shows that ${\mathcal {N}}_m(\rho)\geq \frac{4x-1}{3}$,
which matches with the formula derived in \cite{dnk}. Thus in this
case the lower bound is exact for CREN. Moreover, our lower bound is
experimentally measurable, in the sense that $C_{\alpha\beta}={\rm
Tr}(L_{\alpha}\otimes L_{\beta}\,\rho^{T_A}L_{\alpha}\otimes
L_{\beta})={\rm Tr}(L_{\alpha}\otimes L_{\beta}\,\rho
L_{\alpha}\otimes L_{\beta})$ is the mean value of the Hermitian
operator $L_{\alpha}L_{\alpha}^\dag\otimes L_{\beta}L_{\beta}^\dag$,
and $X(\rho_{\alpha\beta})=\min\{0,d(\rho_{\alpha\beta})\}$ is
determined by the mean value of the operator ${\mathcal
{B}}^{'}_{\alpha\beta}$. On the other hand, according to the proof
of the theorem the lower bound (\ref{lob}) for pure bipartite
quantum states is also exact. Thus based on the continuity of the
CREN, for weakly mixed quantum state $\rho$ with ${\rm
Tr}\{\rho^2\}\approx 1$, (\ref{lob}) supplies a good estimation of
the CREN.

In conclusion, we have derived a set of inequalities
that can detect better entanglement of quantum mixed states. These
inequalities serve as sufficient and necessary conditions
for separability for all bipartite pure states and the isotropic
states. Nevertheless, generally bound entangled states can not be
detected by these inequalities. We also find that these inequalities
have close relations with the convex-roof extension of the
negativity. A measurable lower bound for the convex-roof extension
of the negativity has been obtained.

\bigskip
\noindent{\bf Acknowledgments}\, This work is supported by the NSFC
10875081, NSFC 11105226, KZ200810028013, PHR201007107, the open fund
of State Key Laboratory of Information Security (Graduate University
of Chinese Academy of Sciences)and the Natural Science Fund of
Shandong Province (No.ZR2010FM017).

\end{document}